\documentclass{ws-mplb}
\begin{document}
\markboth{Zhenyue Zhu and Steven R. White}{Quantum phases of the frustrated \textit{XY} models on the honeycomb lattice}

%
\catchline{}{}{}{}{}
%

\title{Quantum phases of the frustrated \textit{XY} models on the honeycomb lattice}

\author{\footnotesize Zhenyue Zhu}
\address{Department of Physics and Astronomy, University of California, Irvine\\
Irvine, CA 92617, USA\\
zhenyuez@uci.edu}

\author{Steven R. White}
\address{Department of Physics and Astronomy, University of California, Irvine\\
Irvine, CA 92617, USA\\
srw@uci.edu}

\maketitle

\begin{history}
\received{(22 October 2014)}
\accepted{(3 November 2014)}
\end{history}

\begin{abstract}
Searching for spin liquid states has long been attracting both
experimentalists and theorists. In this article, we review recent
density matrix renormalization group studies of the spin-1/2
\textit{XY} model on the honeycomb lattice, with first-neighbor
($J_1 = 1$) and frustrating second-neighbor ($J_2>0$) interactions.
For the intermediate frustration regime $0.22\lesssim
J_2\lesssim0.36$, there exists a surprising antiferromagnetic Ising
phase, with ordered moments pointing along the $z$ axis, despite the
absence of any $S_z S_z$ interactions in the Hamiltonian.
Surrounding this phase as a function of $J_2$ are antiferromagnetic
phases with the moments pointing in the $\textit{x-y}$ plane for
small $J_2$ and a close competition between an $\textit{x-y}$ plane
magnetic collinear phase and a dimer phase for large values of
$J_2$. No spin-liquid phases was found in the \textit{XY} model even
with the third neighbor ($J_3>0$) interactions.
\end{abstract}

\keywords{frustration; \textit{XY} model; spin liquid; honeycomb lattice; DMRG.}

\section{Introduction}
Progress in finding model quantum Hamiltonians with spin-liquid (SL)
ground states has accelerated dramatically in the last two years,
almost 40 years since Anderson first proposed a resonating valence
bond (RVB) state as a possible ground state of the triangular
Heisenberg model.\cite{ARVB} However, it was later shown that the
ground state has antiferromagnetic $\sqrt{3} \times \sqrt{3}$ order,
where the moments lie in the same plane with $120^\circ$ angles
between neighboring spins.\cite{tri0}

The main defining feature of a quantum spin liquid is the absence of
any spontaneously broken symmetry, particularly either magnetic or
valence-bond order. Frustration, which discourages order, is a key
ingredient of models potentially containing spin liquid phases. Spin
liquids arise in several analytic treatments and exactly solvable,
simplified, but less realistic models.\cite{SL} A key feature
distinguishing types of spin liquids is the presence or absence of a
gap to all excitations.  To satisfy the Lieb-Schultz-Mattis theorem,
gapped spin liquids for models with a net half-integer spin per unit
cell must have ``hidden'' topological degeneracies in the
thermodynamic limit, which depending on the topology of the system.
These topological SL states possess ``hidden" non-local order ---
long range entanglement. The simplest possibility is a $Z_2$ spin
liquid. There are two classes of lattice models with hard local
constraints, namely the quantum dimer\cite{qdm1,qdm2} and loop
models (toric code\cite {tori} and string-net model\cite{lwen}) that
possess $Z_2$ topological SL ground state.

The quantum dimer model was first introduced by Rokhsar and
Kivelson\cite{qdm1} on the square lattice, where the degrees of
freedom consist of dimers on links with a constraint that there is
only one dimer touching each vertex.  At the RK point, the ground
state is the equal weight superposition of all possible dimer
coverings on the square lattice (RVB state). It was later
generalized by Moessner and Sondhi\cite{qdm2} to the triangular
lattice, where the system has a $Z_2$ topological ground state.
There are four degenerate ground states on a torus, since there are
two non-contractible loops; each loop can possess even or odd
parity, depend on whether a loop cuts an even or odd number of
bonds. Note that on a small finite size torus, these states are not
degenerate, with an energy splitting between different topological
sectors decaying exponentially with the loop size. The toric code
model also has $Z_2$ topological order with 4 fold ground state
degeneracy on a torus.

Since local measurements cannot identify $Z_2$ or other topological
order, it is challenging to identify its presence in a numerical
study.   The degeneracies characteristic of a 2D gapped $Z_2$ spin
liquid have not been accessible for the system sizes studied to
date. Odd-width cylinders spontaneously dimerize in a pattern that
is characteristic of a quasi-one-dimensional system.\cite{kag1}
Besides these properties for a $Z_2$ topological SL state, another
key feature of a $Z_2$ spin liquid is the presence of topological
entanglement entropy (TEE) $\gamma$ introduced by
Kitaev-Preskill\cite{ent1} and Levin-Wen\cite{ent2}. For a
topological phase, $\gamma=-\ln{D}$, where D is the total quantum
dimension of the system. For conventionally ordered phases, D=1. For
topological states, $D>1$. Physically we can understand the origin
of this term using the toric code or string net model. These models
have a wavefunction describing closed loops. Each loop must cross
the boundary twice, thus there is a reduction of entanglement
entropy. For a $Z_2$ topological state $\gamma=-\ln{2}$.

The quantum dimer and toric code models are artificial Hamiltonians
constructed with a delicate topological order. The last few years
have seen a major resurgence in both experimental and theoretical
interest in quantum spin liquid ground states. Much of the interest
stems from strong evidence that quantum spin liquids exist
experimentally in several different materials.\cite{SL} In the case
of the kagome lattice material Herbertsmithite
ZnCu$_3$(OH)$_6$Cl$_2$ with all the Cu$^{2+}$ carrying spins S=1/2
occupy the sites of a Kagome lattice in weakly coupled layers, there
are  substantial experimental evidence that this material has a spin
liquid low temperature phase, with no magnetic or valence bond order
down to 50mk, along with a gapless spin
excitations.\cite{ex1,ex2,ex3,ex4,ex5} The effective Hamiltonian can
be described by the spin S=1/2 Heisenberg Kagome anti-ferromagnetic
model.

This SL state has coincided with recent strong numerical evidence
that the spin-1/2 Heisenberg kagome antiferromagnet has a spin
liquid ground state,\cite{kag1} and that this state has $Z_2$
topological order,\cite{kag2} since a $-\ln 2$ constant term
correction to the linear growth of the entanglement entropy with
subsystem perimeter is observed. The same thing happens for the
kagome system with next-nearest-neighbor interaction
$J_2$,\cite{kag3}
 where for $J_2=0.1$ the gaps are large and the
entanglement entropy correction term can be measured particularly
precisely. Thus, there is now solid evidence that the ground state
of the kagome spin-1/2 antiferromagnet is a gapped $Z_2$ spin
liquid. However, the most important issue is that Herbertsmithite
seems to be a gapless SL, where the $Z_2$ SL state is fully gapped.
Hopefully this discrepancy can be solved by refining the model
Hamiltonian for the real material. So there is great interest in
understanding the kagome SL in more detail, and in finding other SLs
in simple realistic models.

The numerical work has become possible through continued advances in
density matrix renormalization group (DMRG) techniques;\cite
{dmrg1,dmrg2,rev} these methods can now be used to study frustrated
spin Hamiltonians on cylinders with widths up to 12 or 14 lattice
spacings, which, when combined with careful finite size analysis,
can determine phases and properties in the two-dimensional
thermodynamic limit with good confidence in many cases. At the same
time, the reduced density matrix is calculated in the DMRG algorithm
at each sweeping step. Therefore we can directly compute the
entanglement entropy from DMRG.

SL phases have been suggested for various other realistic models,
such as the half-filled honeycomb Fermi-Hubbard model\cite{hubhon}
and the square lattice spin-1/2 Heisenberg antiferromagnet with
second-neighbor ($J_2$) interactions.\cite{squa1,squa2} However,
some skepticism has been expressed about the evidence for spin
liquids in these two models.\cite{hubhon2,jq} There are two very
recent papers pointing out that these two models do not possess SL
ground states. The half-filled honeycomb Fermi-Hubbard model only
has one single phase transition at $\frac{U}{t}\sim 3.78$ between a
semi-metal phase for small U and an AFM state for large
U.\cite{nosl1} A recent DMRG calculation carefully analyzed the
$J_1-J_2$ model on the square lattice and showed that there is a
much stronger square plaquette valence bond pattern for intermediate
coupling at $0.5<J_2<0.6$, showing that the plaquette correlation
length increases faster as the cylinder gets wider.\cite{nosl2}

In 2011, Varney \emph{et al.}\cite{bose} studied the spin-1/2
\textit{XY} model on the honeycomb lattice, with first-neighbor
($J_1=1$, $\langle i,j\rangle$) and frustrating second-neighbor
($J_2 > 0$, $\langle\langle i,j\rangle\rangle$) \textit{XY}
interactions, with Hamiltonian
\begin{equation}
H=J_1\sum_{\langle i,j\rangle}(S_i^+ S_j^- + H.c.)
+J_2\sum_{\langle\langle i,j\rangle\rangle}(S_i^+\ S_j^- + H.c.).
\end{equation}
Based on exact diagonalization (ED) of various small clusters, they
suggested that a particular spin-liquid ground state, a ``Bose
liquid,'' appears for $0.21\lesssim J_2 \lesssim 0.36$. Bose liquids
may have a singular surface in momentum space, similar to a Fermi
surface for a Fermi system, with gapless excitations and power-law
correlations,\cite{bo1,bo2} or they may be gapped and
incompressible.\cite{bo3,mf} This spin model is equivalent to
spinless hard-core bosons with first- and second-neighbor hopping
and zero off-site interactions.

Later on, DMRG calculations on large cylinders showed that although
the locations of the two phase boundaries and the properties of the
phase for small $J_2$ are correct, the intermediate phase has
long-range Ising antiferromagnetic order, instead of the Bose metal
phase.\cite{zhu} This phase was not noticed in previous work on
smaller systems. In terms of bosons, this intermediate Ising phase
has ``charge-density" order of the bosons, with higher density on
one sub-lattice than the other. Afterwards, the coupled cluster
method also verified the existence of this Ising ordered
phase.\cite{cmxy}

In the rest of the paper, we will review the progress of the
$J_1-J_2$ \textit{XY} model on the honeycomb lattice, and also
present some new results unpublished before. In Sec. 2, we present
the results of DMRG calculations on this model, including the
determination of the phase transition points, properties of the
Ising antiferromagnetic order at $J_2=0.3$, and the competition of
the dimer and collinear state at large $J_2$. In Sec. 3, we include
the further neighbor interaction $J_3$ in the Hamiltonian to study
the phase diagram on the XC8 cylinder. In Sec. 4, we study the
transition between $J_1-J_2$ \textit{XY} and Heisenberg model at
$J_2=0.3$. We summarize the results in Sec. 5.

\section{Quantum phases of the $J_1-J_2$ \textit{XY} model}

In this section, we review the results obtained using DMRG on the
$J_1-J_2$ \textit{XY} model on the honeycomb lattice and compare the
results with other numerical techniques, such as series expansions
and the coupled cluster method. In the unfrustrated limit $J_2=0$,
the ground state has the expected N$\acute{\text{e}}$el order in the
$xy$ plane. We find that this phase extends to $J_2\sim 0.22$. In
the interval $0.22\lesssim J_2\lesssim 0.36$, we find an
antiferromagnetic phase that surprisingly has staggered
magnetization polarized along the $z$-direction in spin space; we
call this Ising antiferromagnetic order, to distinguish it from Neel
order in the $xy$ plane. Finally, for $J_2\gtrsim 0.36$, we find
that there is a close competition between a magnetically ordered
$xy$-plane collinear phase and a magnetically disordered dimer
phase.

\subsection{Setup of cylinder geometries for the DMRG calculation}

We have performed numerous DMRG\cite{dmrg1,dmrg2,rev} calculations
on this model on long cylinders with circumferences up to 12 lattice
spacings.  The properties of the ground state are governed by the
ratio $J_2/J_1$. In all of our calculations, we take $J_1=1$ and
$0\leq J_2\leq 1$, thus antiferromagnetic interactions.  The
cylinder geometries we used in our DMRG calculations are adopted
from Ref. \refcite{cyc}. For example, XC8 represents a cylinder
where one set of edges of each hexagon lie along the $x$ direction
(which always coincides with the cylinder axis), and there are 8
spins along the circumferential zigzag columns, connected
periodically (Fig. \ref{phase}(a)). The actual circumference
(Euclidean distance) of XC8 is $C = 4\sqrt{3}$ lattice spacings. For
the YC6 cylinder, one set of edges of each hexagon lies along the
$y$ (circumferential) direction and there are 6 spins (in 3 pairs)
along a straight circumference of $C = 9$ lattice spacings [Fig.
\ref{4-phase3}(b)]. For narrow cylinders like XC8, we are easily
able to achieve a truncation error of about $10^{-8}$ with M = 2400
states, which determines the ground state essentially exactly. For
YC8, our widest cylinder with $C=12$, we need to keep M = 5800 to
achieve a truncation error of $10^{-6}$---still excellent accuracy.
In all our DMRG calculations, we keep enough states to make sure
that the truncation error is smaller than $10^{-6}$.

When performing DMRG calculations on cylinders, we have to map the
lattice into a one dimensional chain and start the regular sweeping
process. Let A and B denote two different sub-lattice on the
honeycomb lattice. In the XC cylinders, we find that it is better to
map into the 1D vertical zigzag chains one by one (AB site chains
.....), instead of all the straight columns of sites (A site chain,
B site chain .....). Otherwise, it is easy to get stuck in a higher
energy state on wider cylinders which have magnetic domain walls .

\subsection{Determination of the phase transitions}

The classical phase diagram of the $J_1-J_2$ \textit{XY} model on
the honeycomb lattice has the following phase diagram: for
$0<J_2<\frac{1}{6}$, it is the Neel order with wave vector Q located
at the center of the Brillouin zone ($\Gamma$ point);  For
$\frac{1}{6}<J_2<\frac{1}{2}$, it is the coplanar spiral I phase
with Q forms closed contours around the $\Gamma$ point; At
$J_2=\frac{1}{2}$, it is the collinear state with nearest neighbor
bonds are ferromagnetic in one direction and antiferromagnetic in
the other two direction with the closed contour has a hexagonal
shape and touch the edge center of the Brillouin zone (M point); For
$J_2>\frac{1}{2}$, it is the spiral II phase with closed contours
around the corner of the Brillouin zone (K or K' point).\cite{cla}
Considering quantum correlations to the classical spin waves, it is
found that the Neel state extends up to $J_2=0.26$, well beyond the
classical value $\frac{1}{6}$. On the other hand, the collinear
state is stable from $J_2=\frac{1}{2}$ to a regime of
$0.26<J_2<1$.\cite{cla}  In the quantum limit of the phase diagram,
the critical point usually expands to an the intermediate phase.
Therefore, we would expect that there will be a intermediate phase
between Neel order and collinear order state. Next we will apply
DMRG to investigate the quantum phase diagram.

\begin{figure}[htcp]
\begin{center}
\includegraphics*[width=11cm, angle=0]{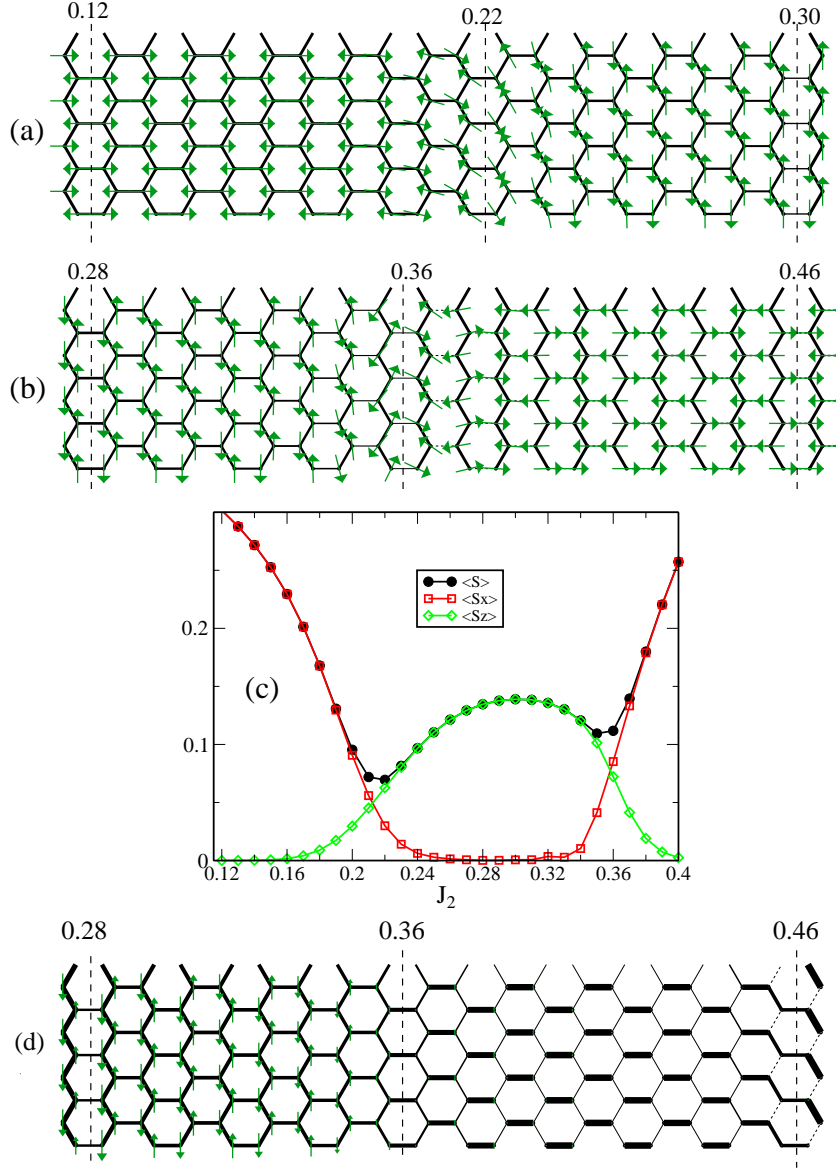}
\caption{ (a) Local magnetic moments along the XC8-0 cylinder with
$J_2$ varying along the length of the cylinder. The length of the
arrows are fixed, while their angle correctly represents the spin
orientation. The dashed lines show particular $J_2$s, which are
constant along a column of hexagons. In (a), we show a system with
$J_2$ varying from 0.12 to 0.30. For $J_2\sim 0.22$, the xy-plane
Neel order flips to z-direction Ising order. In (b), $J_2$ is varied
from 0.28 to 0.46. The second phase transition point is located at
$J_2\sim0.36$ between Ising and metastable collinear state. (c)
shows the size of $|\langle S\rangle|$, $|\langle S_x\rangle|$,
$|\langle S_z\rangle|$ along the cylinder column versus $J_2$. The
phase transition is apparent at $J_2$s where $|\langle S\rangle|$ is
a minimum. (d) shows the phase transition from Ising order to dimer
order with phase transition point is also located at $J_2\sim0.36$.
}\label{phase}
\end{center}
\end{figure}

In Fig. \ref{phase}, we present two cylinders to first give a quick
summary of the whole phase diagram. These are XC8 cylinders in which
$J_2$ is varied along the length of the cylinder, showing locally
the various phases. In Fig. \ref{phase}(a), $J_2$ varies from 0.12
to 0.30. At the $J_2=0.12$ left edge, a staggered field in the $xy$
plane was applied to ``pin'' the Neel order. The ordered moments
rotate from the $x$ to the $z$ direction, indicating that there is a
phase transition between Neel and Ising order, at $J_2\sim0.22$. In
Fig. \ref{phase}(b), $J_2$ is varied from 0.28 to 0.46, with AF
pinning fields along the x direction at the $J_2=0.46$ right end to
show the collinear pattern. The phase transition from Ising to
metastable collinear order is visible at $J_2\sim0.36$, where spin
flips to $x$ direction again. In Fig. \ref{phase}(c), we show the
actual size of $|\langle S\rangle|$, $|\langle S_x\rangle|$,
$|\langle S_z\rangle|$ for each $J_2$. It's clearly seen that the
intermediate phase has $|\langle S_z\rangle|\sim0.14$ with $|\langle
S_x\rangle|=0$. The minimum of $|\langle S\rangle|$ locates the two
phase transition points at $J_2=0.22$ and 0.36. We also show in Fig.
\ref{phase}(d) that the phase transition between Ising order and
dimer order is also located at $J_2=0.36$, since ground state in the
XC8-0 cylinder at large $J_2$ is the a dimer instead of collinear
state. See next section for the detailed discussion of the collinear
and dimer states. We also applied these methods to other cylinders
and find that the values of $J_2$ at the estimated phase transitions
change only slightly between different width and orientation
cylinders. Thus the locations of these phase transitions show only
small finite size effects, which is consistent with our agreement
with the small-size ED results from Ref. \refcite{bose}.

\begin{figure}
\centerline{\psfig{file=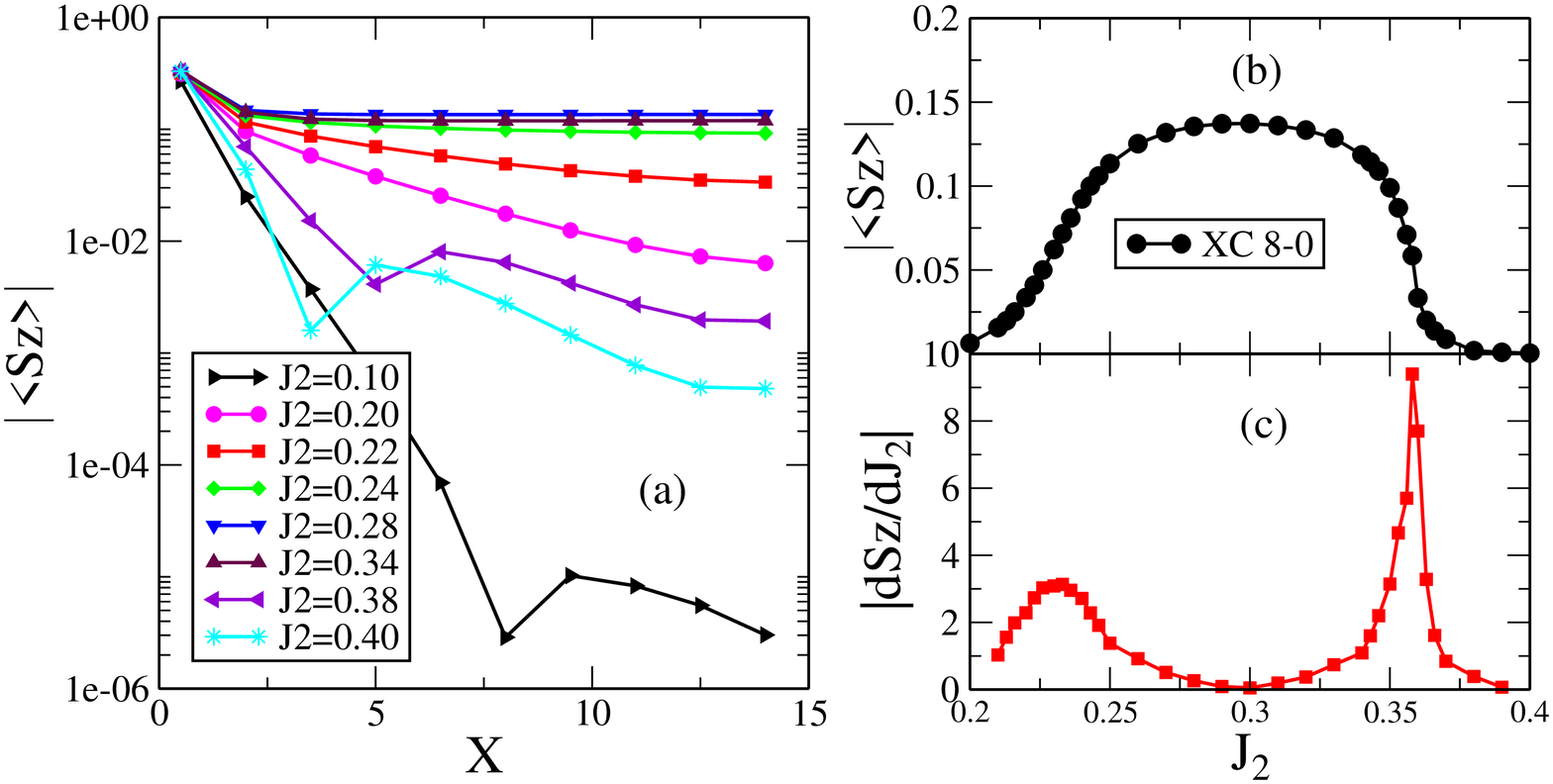, width=8.5cm}} \caption{(a) Local
magnetization $|\langle S_z\rangle|$ versus distance from the end of
an XC8 cylinder for various $J_2$ values. (b) The magnetization at
the cylinder center versus $J_2$. (c) The derivative of the central
$|\langle S_z\rangle|$ versus $J_2$. The peaks of the derivative at
$J_2=0.23$ and $0.36$ indicate the two phase transition points. From
Zhu \emph{et al.} Ref. \protect\refcite{zhu}. }\label{psz}
\end{figure}

In Fig. \ref{psz}, we apply a staggered field with $h_z=0.5$ at a
cylinder end to measure the decay of $|\langle S_z\rangle|$ with
distance from the end for various values of $J_2$. As shown in Fig.
\ref{psz}(a),  $|\langle S_z\rangle|$ decays exponentially within
both the Neel and the dimer phases, but the correlation length gets
longer and $|\langle S_z\rangle|$ becomes spatially uniform in the
cylinder center for the Ising ordered phase. In Fig. \ref{psz}(b),
we show the magnetization in the cylinder center versus $J_2$.  It
is clear from this plot that the intermediate Ising phase is a broad
regime, and from its derivative versus $J_2$, we determine the two
phase transition points at $\sim 0.23$ and 0.36, which approximately
match the phase transitions determined from Fig. \ref{phase}.  It is
interesting to note that $|\langle S_z\rangle|$ is almost
independent of $J_2$ for much of this intermediate Ising phase.  The
moment $|\langle S_z\rangle| \sim 0.14$ is strongly reduced from the
maximum ``classical'' value of 0.5.

The derivative of $|\langle S_z\rangle|$ shown in Fig. \ref{psz}(c)
shows markedly different behavior for the two transitions, with the
second transition being much sharper. A natural interpretation is
that the phase transition between Neel and Ising phases is
continuous, but the second phase transition point is first order. To
test this, we have performed calculations on cylinders with a much
narrower range of $J_2$ values along the length of the cylinder,
zooming in on the transitions. If the phase transition is first
order, we expect that the phase transition region should remain
narrow as we zoom in. For a continuous phase transition, the phase
transition region should broaden as we zoom in. Varying the gradient
of $J_2$ by a factor of 5, we do find that the Neel-Ising phase
transition region broadens, but the second phase transition region
stays narrow.  Thus, it does appear that the former is continuous,
and the latter is first order. However, any conclusions about the
second transition are tentative, because of the close competition
between the dimer and collinear phases for $J_2>0.36$.

In comparison, the fidelity measurement from the exact
diagonalization on small size clusters found two phase transition
points at $0.210(8)$ and $0.356(9)$.\cite{bose} Series expansions,
measuring the local magnetization in the Neel and collinear phases,
determined phase transitions at $0.22(1)$ and around
$0.35$.\cite{sexy} Similarly, the coupled cluster method determined
the two phase transition points at $0.216(5)$ and
$0.355(5)$.\cite{cmxy} Therefore, it is believed that the two phase
transition points are located around $0.22$ and $0.36$ in the
thermodynamic limit with an intermediate phase between them.

\subsection{The detailed study of ground state properties at $J_2=0.3$}

We have tested the stability of the Ising phase in several ways. For
example, one can measure the decay of the local staggered
magnetization away from an applied staggered field on an end of the
cylinder. For the Neel ordered phase (small $J_2$), when we apply
the pinning magnetic field along the $z$ direction, $|\langle
S_z\rangle|$ decays exponentially from the cylinder end [Fig. 3(a)].
To similarly test the Ising phase, we apply the pinning field along
the $x$ direction at the ends of an XC8 cylinder with $J_2=0.3$.  We
find that $|\langle S_x\rangle|$ decays exponentially with distance
from the cylinder end with a very short correlation length
$\xi_x=1.8$, but $|\langle S_z\rangle|$ rises from the end and
saturates in the center of cylinder (not shown). This provides solid
evidence that Ising order is very robust on this cylinder. As
another test, we have measured the correlation function $|\langle
S_i^+(0) S_j^-(x) \rangle|$ and find that its correlation length
decreases as a function of increasing $J_2$ for $J_2$ near 0.22, and
then increases rapidly for $J_2$ near 0.36. The minimum correlation
length is roughly $\xi\sim1.5$ at $J_2=0.3$ (not shown). This result
again confirms that $xy$-plane order is absent in the intermediate
phase.

\begin{figure}
\begin{center}
\includegraphics*[width=11cm,angle=0]{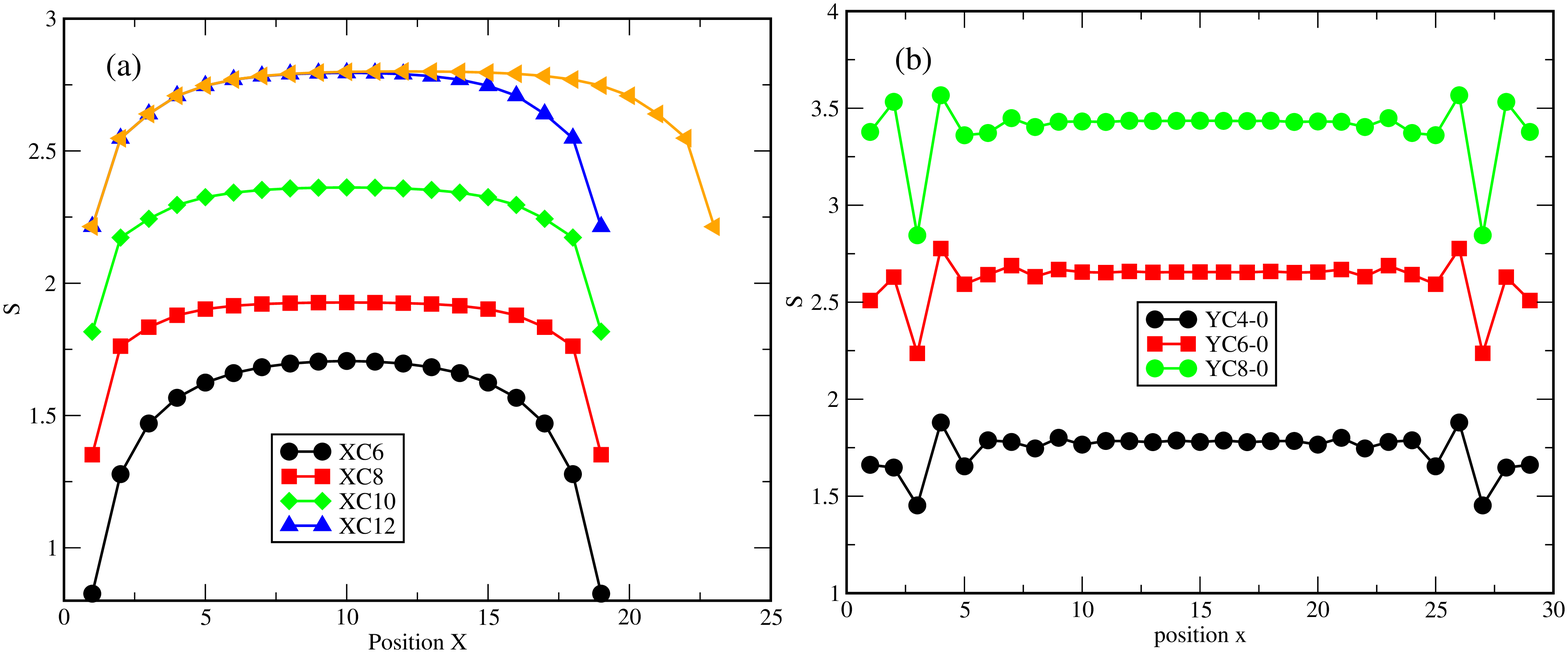}
\caption{The unextrapolated ground state entanglement entropy versus
subsystem size x for (a) XC cylinders and (b) YC cylinders. The
entanglement entropy for the widest cylinder shown here is not quite
converged, where the number of states kept is M=5600. }\label{entro}
\end{center}
\end{figure}

\begin{figure}
\centerline{\psfig{file=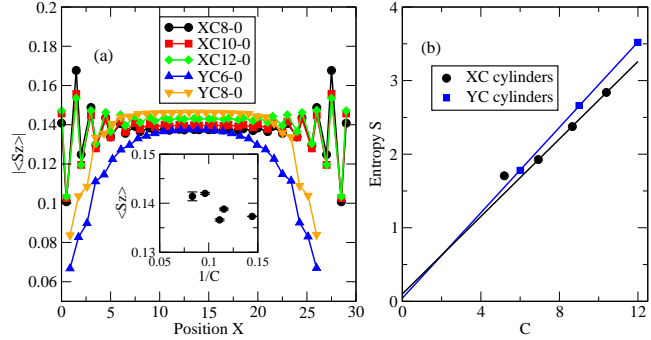, width=8.5cm}} \caption{(a) The
absolute value of the local magnetization $|\langle S_z\rangle|$ for
various XC and YC cylinders versus distance along the cylinder. The
inset shows the extrapolated magnetization (extrapolated versus the
truncation error) at the cylinder center, with error bars, versus
the inverse of cylinder circumference $C$. (b) The entanglement
entropy versus circumference for XC and YC cylinders in the
intermediate Ising ordered phase at $J_2=0.3$. The intercepts are
consistent with zero. From Zhu \emph{et al.} Ref.
\protect\refcite{zhu}. }\label{mag-s}
\end{figure}

We also measured the entanglement entropy for various cylinder sizes
and extrapolated to see if there is a possible topological
entanglement entropy contribution ($\gamma$) in Fig. \ref{mag-s}(b).
Entanglement entropy area law states that for a gapped phase
entanglement entropy only depends on the boundary size, independent
of the subsystem size.
\begin{equation}
S\sim a L+\gamma+O(1/L),
\end{equation}
with L its boundary length.\cite{ea} In Fig. \ref{entro}, we show
the entanglement entropy versus the subsystem size x for various XC
and YC cylinders. For XC cylinders, the entropy is lower at the
cylinder ends, but saturated in the cylinder center independent of
the cylinder length. The entropy oscillates near the YC cylinder
edges, because the system shows plaquette pattern only at the edges
with short PVB correlation length.

We then use the entropy at the center for various cylinders
(extracted from Fig. \ref{entro}) to linear extrapolate $\gamma$
from the above equation. For a $Z_2$ spin liquid, one would expect
$\gamma=-\ln 2$. We find $\gamma \sim 0.09$ for XC cylinders and
$\gamma \sim 0.04$ for YC cylinders, values consistent with zero, as
expected for a non-topologically ordered state. We expect that if we
could include larger cylinders, the resulting data would extrapolate
to $\gamma = 0$.

To make sure that the Ising order is not a finite size effect, we
have studied the $J_2=0.3$ system for cylinders with various widths.
Figure \ref{mag-s}(a) shows $|\langle S_z\rangle|$ as a function of
$x$ for XC and YC cylinders. In the inset, we plot the extrapolated
magnetization at the cylinder center versus the inverse of the
cylinder circumference.  For these cylinders, the staggered
magnetization is nearly constant with circumference, taking a value
of about $0.135-0.142$. Thus, we believe that $|\langle
S_z\rangle|\sim0.14$ in the 2D limit for $J_2=0.3$.  If anything,
the staggered magnetization increases with increasing $C$, so this
should be viewed as a lower bound on the value in the 2D limit.

We have not been able to find a simple analytical argument or
calculation that gives an intuitive picture for this robust Ising
ordered state.  However, viewing the system as hard-core bosons at
half filling provides an additional perspective. The Hamiltonian can
be mapped straightforwardly and exactly into a hard-core boson model
with first-neighbor hopping $t_1 = J_1/4$ and second-neighbor
hopping $t_2 = J_2/4$, since $S^{\dagger}=b^{\dagger}/2$,
$S^z=b^{\dagger}b-0.5$. The Ising order would appear as a charge
density wave (CDW) order, where the density is higher on sub-lattice
\textit{A} ($n_A\sim 0.64$) than on sub-lattice \textit{B}
($n_B=1-n_A\sim 0.36$). Although there are only hopping terms in
this hard-core boson Hamiltonian, the hard-core constraint (one
boson per lattice site) is an on-site interaction.  One could
imagine that this on-site interaction renormalizes in some way to
produce a first-neighbor density-density interaction, which could
produce the CDW. This system is the first that we are aware of where
a CDW is produced only from the combination of frustrated hopping
and a hard-core constraint.

\begin{figure}
\centerline{\psfig{file=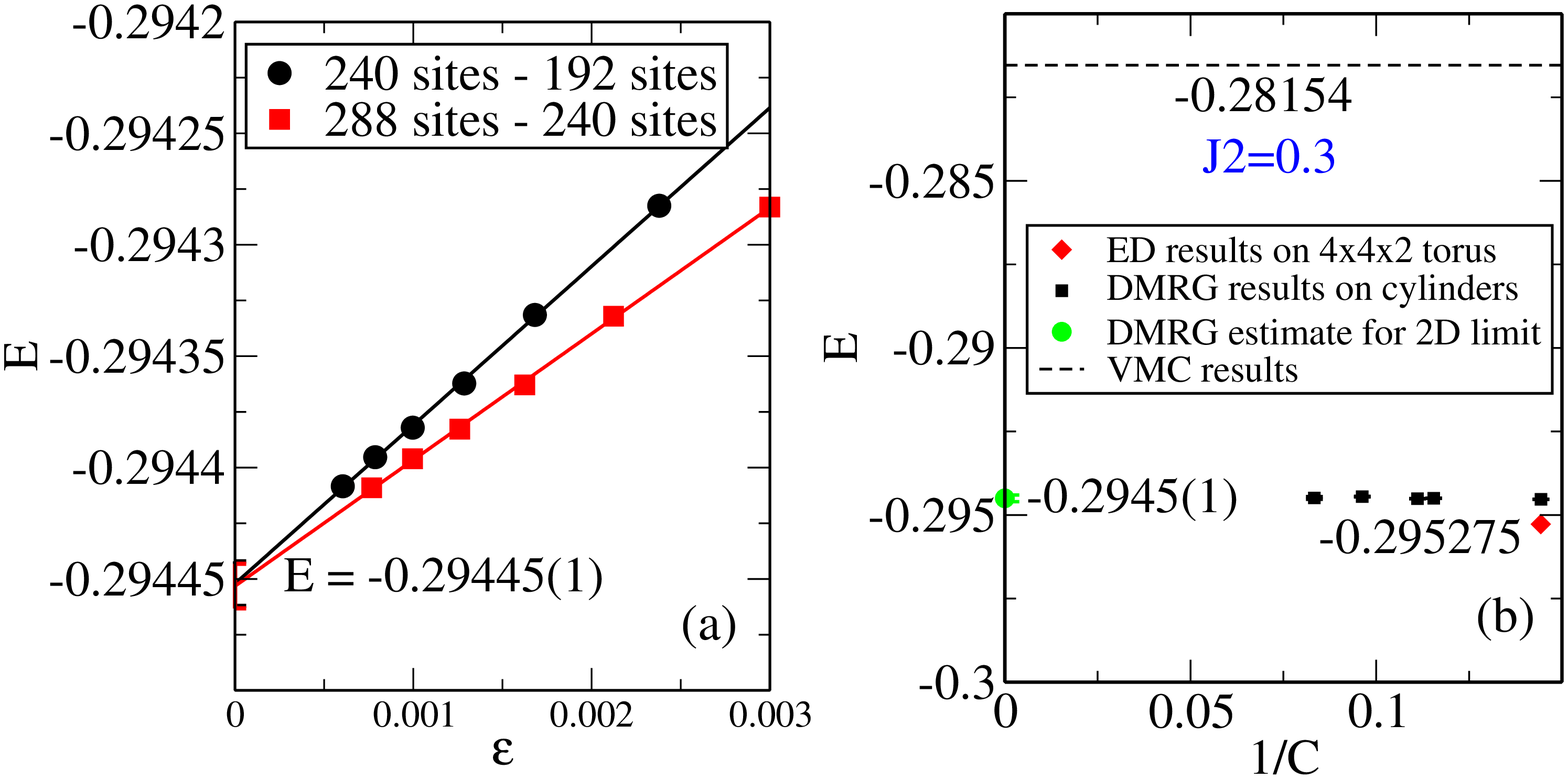,width=8.5cm}} \caption{(a) The
extrapolation of the ground state energy per spin for $J_2=0.3$
versus truncation error for the XC12 cylinder. The black curve is
the energy per spin from subtracting the energies of two XC12
cylinders with lengths $L_x =20 $ and $L_x=16$. The red curve is
from subtracting two cylinders with $L_x =24 $ and $L_x=20$. These
two subtractions extrapolate to the same energy per spin of
$-0.29445(1)$. (b) The ground state energy per spin for $J_2=0.3$
versus the inverse of cylinder circumference from our DMRG
calculations, compared with the variational Monte Carlo (VMC) result
from Table III in the supplemental material, and exact
diagonalization. From Zhu \emph{et al.} Ref. \protect\refcite{zhu}.
} \label{ener}
\end{figure}

Very recently, a variational Monte Carlo study of this model has
appeared.\cite{vmc} In Ref. \refcite{vmc}, a variational spin-liquid
wave function is constructed by decomposing the boson operators into
a pair of fermions with a long-range Jastrow factor, with Gutzwiller
projection enforcing single occupancy. At $J_2=0.3$, the lowest
energy for such a state had energy per spin $E=-0.28154$, which is
$\sim 4\%$ higher than ED of the $4\times4\times2$ torus
($E=-0.295275$)\cite{vmc}. Although this might appear to be a small
difference in energy, for competing phases in geometrically
frustrated spin-1/2 models near spin liquids, this is actually a
very large energy difference.  For example, the spin-1/2 kagome
antiferromagnet (with $J_2=0$) has an energy difference of only
about $1\%$ between the (metastable) honeycomb valence bond crystal
and the spin-liquid ground state.\cite{kag1} In our DMRG
calculations, the energy per spin for a specific cylinder geometry
can be calculated by subtracting two cylinders with the same width
but different lengths.\cite{rev} When the cylinder is long enough,
this method gives the energy per spin in the cylinder center, with
minimal edge effects. We show in Fig. \ref{ener}(a) that the energy
per spin from subtracting two different pairs of cylinders gives
precisely the same energy for the XC12 cylinder. Thus, we find that
the ground state energy per spin is -0.29445(1) for an infinitely
long XC12 cylinder at $J_2=0.3$. In Fig. \ref{ener}(b), we compare
our DMRG results for the ground state energy on various cylinders at
$J_2 = 0.3$. For the cylinders we study, the DMRG energies have
quite small finite size effects. We estimate that the ground state
energy is $E=-0.2945(1)$ in the 2D limit. The small-size ED result
is only slightly ($\sim 0.26\%$) lower in energy, due to its finite
size effects.  The state DMRG finds has antiferromagnetic Ising
order with the spin moments ordered in the $z$ direction. This
ordered ground state has much lower energy than the variational
spin-liquid state.

At $J_2=0.3$, series expansions fail to give accurate ground states
for the Ising phase due to poor convergence. However, indirect
evidence to support the Ising antiferromagnetic order phase is that
the $\langle S_x^iS_x^j+S_y^iS_y^j \rangle$ correlations decrease,
while the $\langle S_z^iS_z^j\rangle$ correlations increase with
$J_2$ and approach the same value near the first phase transition
point at $J_2=0.22$.\cite{sexy} It's expected that the $\langle
S_z^iS_z^j\rangle$ correlations will dominate in the intermediate
phase regime as in the z-direction Ising AF order. The coupled
cluster method starts from the Ising ordered state as a reference
state and finds a ground state energy $E=-0.2947$ and magnetization
$M=0.138$ in the 2D limit, very similar to the DMRG
results.\cite{cmxy} It would interesting to check these results with
other numerical techniques in the future.

\subsection{The competition between VBC and collinear order for large $J_2$}

\begin{figure}
\begin{center}
\includegraphics*[width=8.5cm, angle=0]{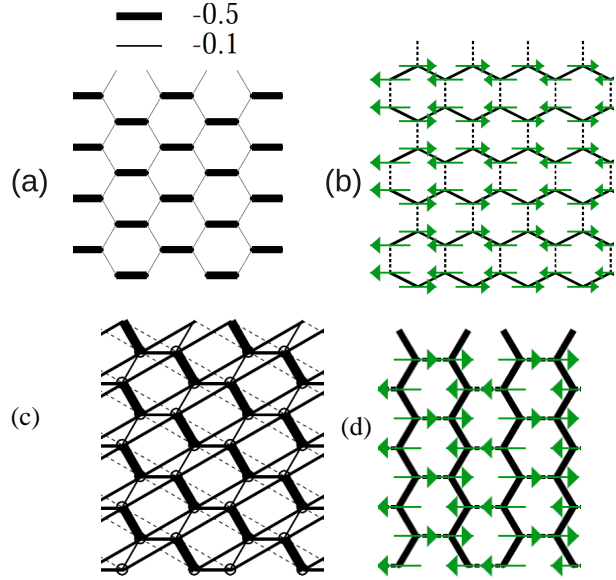}
\caption{(a) The ground state on the XC8 cylinder at $J_2=0.5$,
showing strong dimer correlations aligned horizontally. We call this
state a dimer state or SVBC. It has strong $J_2$ bonds connecting
these dimers vertically and wrapping around the cylinder (not
plotted in this figure). (b) The ground state on YC6 cylinder at
$J_2=0.5$. It has $xy$-plane collinear magnetic order, with
antiferromagnetic chains along the horizontal zigzag direction
together with ferromagnetic first-neighbor correlations between
these zigzag chains. (c) The ground state on XC8 cylinder at
$J_2=0.37$ shows the diagonal SVBC with ladder pattern. (d) The
metastable collinear state on the XC8-0 cylinder at $J_2=0.5$.
}\label{4-phase3}
\end{center}
\end{figure}

At large $J_2$s, we find that both collinear and dimers state can be
stabilized on even XC cylinders. We will start from the XC8-0
cylinders and discuss how we analyze this issue. When we perform
DMRG calculations from different random states, we can produce
different final states. For example, at $J_2=0.4$, we can easily get
a collinear state (Fig. \ref{4-phase3}d) and (less likely) also get
a dimer state (Fig. \ref{4-phase3}a). The opposite happens for
$J_2=0.5$. Therefore it appears that both dimer and collinear states
are stable on the XC8-0 cylinder. We add extra terms in the
Hamiltonian to prepare different initial state and then remove these
extra terms after several sweeps. For example, to start from a dimer
state, we strengthen the horizontal bonds ($S_i^+ S_j^-+h.c.$) of
the XC cylinder by $20\%$. For an initial collinear state, we weaken
these bonds by $20\%$ to enforce the strong bonds along the vertical
zigzag direction. Since the collinear state has magnetic order along
the xy plane, we can also apply the pinning field along the x
direction on the cylinder edges and measure the $S_x$ locally, which
is possible by not keeping the quantum number $S_z$ conserved.
Starting from different possible collinear states on the XC8-0
cylinder, it turns out that only the collinear pattern shown in Fig.
\ref{4-phase3}(d) is a stable collinear state. These results are
presented in the following table. The second and fourth column are
calculated with quantum number $S_z$ conserved, while the third
column is calculated without quantum number conservation, but with
fields applied along the x direction on the cylinder edges to show
the proper collinear magnetic pattern. The two different
calculations for the collinear state have exactly the same energy,
but the ground state for these cylinders are actually a dimer state.

\begin{table}[htp]
\tbl{The energy for different states on the XC8-0 cylinder at various $J_2$s.
}
{\begin{tabular} {|c|l|l|l|}
 \hline
$J_2$       & collinear       & edge $h_x$    & dimer      \\
\hline
0.40        & -0.29815        & -0.29814      &  \textbf{-0.3007}    \\
\hline
0.45        & -0.30778        & -0.30778      & \textbf{-0.3112}    \\
\hline
0.50        & -0.31896        & -0.31895      & \textbf{-0.3229}    \\
\hline
\end{tabular}}
\end{table}

\begin{figure}
\begin{center}
\includegraphics*[width=10cm,angle=0]{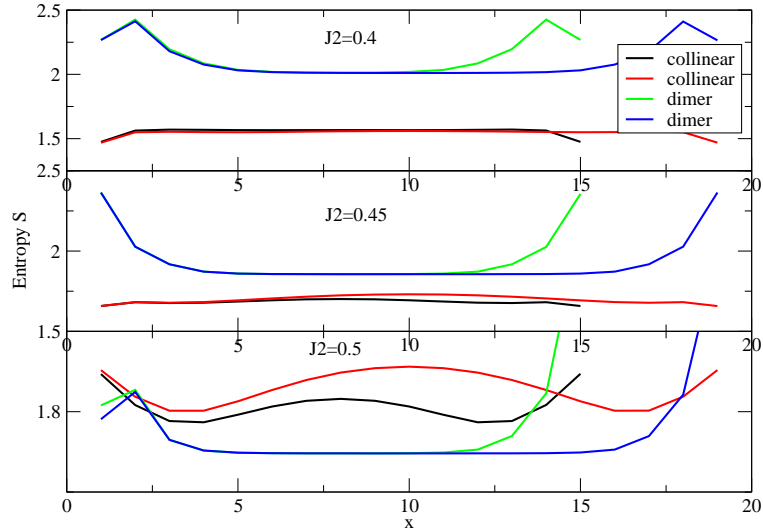}
\caption{The entanglement entropy for collinear and dimer states at
different $J_2$s versus the position of the entanglement cut for the
XC8-0 cylinders. There are two collinear and two dimer states with
different length for each subfigure. }\label{4-s-l}
\end{center}
\end{figure}

We look at the entanglement entropy to understand these different
states in Fig. \ref{4-s-l}. At $J_2=0.4$, the entanglement entropy
for the collinear state is much lower than the dimer state. Since we
know that the DMRG algorithm prefers the low entropy state if
energies are nearly degenerate, DMRG could easily get the collinear
state (low entropy but high energy state) from a random state at
$J_2=0.4$, instead of the dimer state (high entropy but low energy
state). At $J_2=0.5$, the dimer state actually has lower entropy
than the collinear. Thus DRMG finds the correct ground state - a
dimer state with low entropy and low energy. The entropy is measured
from dimer and collinear states with quantum number $S_z$ conserved.
For the collinear state with edge fields applied along the $x$
direction, the entropy is lower, since this state has only magnetic
moments pointing in one specific direction, instead of a
superposition of all possible directions in the $xy$ plane.

The horizontal dimer state on the XC8-0 cylinder has strong $J_2$
bonds connecting them to form ladders wrapping periodically around
the cylinder with length 4. Therefore the horizontal dimer state can
only appear on even XC4N-0 cylinders with ladder length 2N to
accommodate the AFM correlation. Thus we do not obtain stable
horizontal dimer states on the XC10-0 cylinder. For the YC
cylinders, the possible dimer state should be vertical with $J_2$
bonds connecting them to form the ladders horizontally. However, we
find that at $J_2=0.5$, vertical dimer states can not be stablized
in our DMRG calculation. Initially pinned dimer states always evolve
to collinear states. The only stable dimer state on the YC cylinder
is at small $J_2$, say $J_2=0.4$. Another interesting thing about YC
cylinders is that the collinear state pattern can have AFM
correlation on the zigzag chains either along the horizontal
direction (Fig. \ref{4-phase3}(b)) or wrapping around the cylinder
with some angle (when we start from a random state). These two
states have very similar energies. The results are presented in the
following table for $J_2=0.40$, 0.45 and 0.50.

\begin{table}[htp]
\tbl{The energy for different states on various cylinders at
different $J_2$s. The ``N/A" symbol means that the particular state
can not be stabilized on that cylinder. } {\begin{tabular}
{|c|c|c|c|c|c|c|}
 \hline
      &\multicolumn{2}{ c| }{$J_2=0.40$}  &\multicolumn{2}{ c| }{$J_2=0.45$} &\multicolumn{2}{ c| }{$J_2=0.50$} \\
\hline
          & Dimer & Collinear     & Dimer & Collinear     & Dimer    & Collinear \\
\hline
XC8-0     & \textbf{-0.3007}   & -0.2981   & \textbf{-0.3112}   & -0.3078   & \textbf{-0.3229}   & -0.3189   \\
\hline
XC10-0    & N/A         & -0.2981   & N/A         & -0.3077   & N/A         & -0.3188   \\
\hline
XC12-0    & \textbf{-0.2985}   & -0.2978   & \textbf{-0.3083}   & -0.3076   & \textbf{-0.3196}   & -0.3188   \\
\hline
YC4-0     & \textbf{-0.2982}   & N/A         & N/A         & -0.3078   & N/A         & -0.3189   \\
\hline
YC6-0     &-0.2978    & -0.2978   & N/A         &-0.3070    & N/A         & -0.3189    \\
\hline
\end{tabular}}
\end{table}

In this table, we can see that the collinear states on these
cylinders have relatively small finite size effects at fixed $J_2$s.
The collinear states can be stabilized on all the cylinders at
different $J_2$s, while the dimer state can be stabilized on XC8-0
and XC12-0 cylinders, where they are the real ground states. On YC
cylinders, dimer states can only be obtained for small $J_2$. The
other tendency is that the energy difference between dimer and
collinear state decreases as the cylinder becomes wider. At
$J_2=0.5$, the dimer state is $1.2\%$ lower than the collinear state
on the XC8-0 cylinder, as compared to $0.25\%$ lower on the XC12
cylinders. It may be possible that the ground state is the collinear
state in the 2D limit for larger $J_2$s.

Another interesting feature is that the diagonal dimer state is the
ground state on all the cylinders for $J_2$ close to the second
phase transition point ($J_2=0.37$). Therefore, we suspect that
there might be a small phase region for $0.36<J_2<0.4$ where the
ground state is the dimer state. At larger $J_2$s, the collinear
state is more likely to be the ground state.  In Ref. \refcite{vmc},
the authors found that the energy difference between the VMC
collinear state and small size exact diagonalization has a
relatively large error of about $2.5\%$ near the second phase
transition point $J_2=0.35$ and a constant error of about $2\%$ at
$J_2>0.4$. This means that the variational wavefunction for the
collinear state still doesn't quite capture the properties of the
ground state (the dimer state) near $J_2=0.35$.

Series expansions find that the dimer state is marginally lower than
the collinear state at $J_2=0.4$, while the collinear state is
slightly lower at $J_2=0.5$. Therefore at large $J_2$, there will be
another phase transition point between the dimer and collinear
state.\cite{sexy} The coupled cluster method for the dimer response
in the collinear state finds a strong competition between collinear
and dimer states at $0.355(5)<J_2<0.52(3)$. At $J_2>0.52(3)$, the
collinear state is more favorable.\cite{cmxy} Although the dimer
state first found by DMRG was not predicted from either small size
exact diagonalization or VMC calculations,\cite{bose,vmc} all these
numerical methods both agree that there is a close competition
between dimer and collinear state closer to $J_2=0.36$, albeit it is
delicate to determine the transition between dimer and collinear
state.

\section{Phase diagram of $J_1-J_2-J_3$ \textit{XY} model on the XC8-0 cylinder}

\begin{figure}
\begin{center}
\includegraphics*[width=11cm,angle=0]{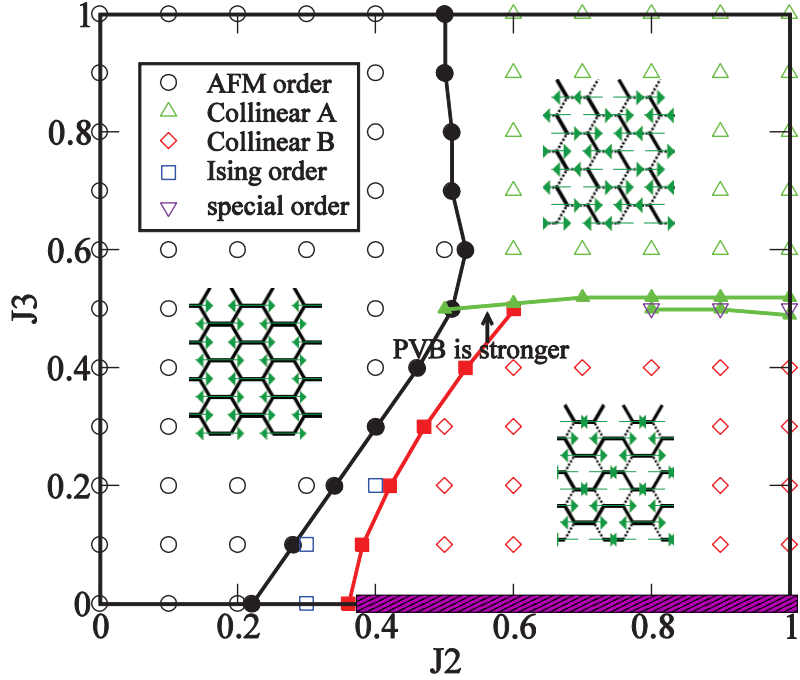}
\caption{The phase diagram with $J_1-J_2-J_3$ \textit{XY}
interactions on the honeycomb lattice. AFM and two types of
collinear orders have moments in the $xy$ plane. The Ising moments
are along the $z$ direction. Each symbol represents the state
calculated with fixed $J_2$ and $J_3$. The solid lines are the
approximate phase transition points from $J_2$ or $J_3$ variation
calculations on a single cylinder. The insets show what the three
main phases (AFM and collinear A/B phase) look like, with solid bond
indicating that nearest neighbor $\langle S_i\cdot S_j\rangle$ is
negative and dashed bond for positive. See the text for details on
the special order phase. The purple pattern at the $J_3=0$ axis with
$J_2>0.36$ is the region where the ground state is the dimer state
on the XC8-0 cylinder. We do not understand how this phase is
connected with the collinear B phase. }\label{4-j3}
\end{center}
\end{figure}

Searching for possible SL states in the honeycomb lattice with only
\textit{XY} interactions, we added third-neighbor interactions
($J_3$ terms) to the system to scan the whole phase diagram in Fig.
\ref{4-j3}. Since adding another parameter $J_3$ greatly increase
the numerical effort, we only performed DMRG calculations on XC8-0
cylinders. We first calculated the ground states on a set of XC8-0
cylinders with $J_2$ and $J_3$ starting at $J_2=0$ and $J_3=0$, then
increase each parameter separately by 0.1 on each cylinder. The
couplings were constant along the length. By applying pinning
magnetic fields along the $x$-direction on the cylinder edges, we
characterized each specific state. The phase transition lines are
determined by calculating a cylinder with $J_2$ or $J_3$ varying
along the length while fixing the other parameter as in Fig.
\ref{phase}(a).

$J_3$ interaction actually favors Neel order. Therefore, we expect
that the phase transition point between Neel and Ising order should
move to higher $J_2$ with increasing $J_3$. This is confirmed on the
bottom solid black line on Fig. \ref{4-j3} for $J_3<0.5$. Then with
$J_3>0.5$, there is a phase transition between Neel and collinear A
ordered phase. Collinear A phase only have AFM spins along one
particular diagonal direction. The phase transition between them is
located roughly at $J_2=0.5$ and $J_3>0.5$. This phase boundary
matches the classical phase diagram for Neel and collinear A phase.

The red solid line is the phase transition between Ising and a
different collinear B phase. The intermediate phase region (between
solid black and red line) has very robust Ising order at small
$J_3$. But Ising order parameter decreases as $J_3$ increasing. At
around $J_2=0.6$ and $J_3=0.5$, valence bond order is much stronger.
Therefore, we suspect that there might be a phase transition between
Ising and valence bond order for the intermediate phase regime. We
will not focus on this issue in the paper.

The green line is the phase boundary between collinear A and B
phase. At $J_2$ close to 1, we find that there exists a small region
at $0.49<J_3<0.52$ and $0.8<J_2<1.0$, which we call the special
order phase. This phase has very weak nearest neighbor bonds, but
with strong bonds along the vertical second-neighbor and horizontal
third-neighbor directions forming a rectangular lattice. This
pattern breaks the honeycomb lattice six fold rotational symmetry,
and thus it is also not a spin liquid state. It would be interesting
to check if this phase is presented on the wider cylinders.

Note that the collinear B phase is different from the collinear
phase in Fig. \ref{4-phase3}(d) at $J_3=0$. At $J_3=0$, the
metastable collinear phase has AFM correlated zigzag chains with FM
correlations between them. But with a little $J_3$ interaction added
to the system, the ground state forms a different type of collinear
phase with AFM correlated chains along the armchair direction of the
hexagon lattice. It's hard for us to determine the phase transition
between these two collinear states specifically. But we can
understand this from the following classical picture, instead of
including classical spin spiral states.

The classical energy for a Neel state $E_N$, for a collinear state
with AFM correlation in zigzag chains (collinear Z state in Fig.
\ref{4-phase3}d) $E_z$, for a collinear A state $E_A$ and for a
collinear B states $E_B$ are as follows:
\begin{eqnarray}
E_{N} =& -\frac{3}{8}&(J_1-2J_2+J_3)\nonumber \\
E_{z} =& -\frac{1}{8}&(J_1+2J_2-3J_3)\nonumber \\
E_{A} =& \frac{1}{8}&(J_1-2J_2-3J_3)\nonumber \\
E_{B} =& -\frac{1}{8}&(J_1+2J_2-J_3)
\label{enere}
\end{eqnarray}

\begin{figure}[htbp]
\begin{center}
\includegraphics*[width=8cm,angle=0]{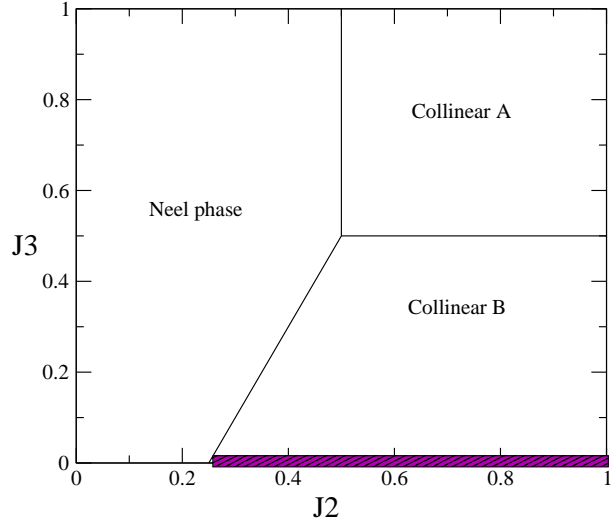}
\caption{The classical phase diagram for the $J_1-J_2-J_3$
\textit{XY} model, where we only consider several classical states
-Neel, collinear Z, A and B states. Only at $J_3=0$ and $J_2>0.25$
(the purple region), collinear Z and B states have the same energy.
}\label{4-class}
\end{center}
\end{figure}

For $J_3=0$, two classical collinear states $E_z$ and $E_B$ have
exactly the same energy. However with $J_3>0$, only the classical
collinear B state has lower energy. These two collinear states in
the quantum limit have similar properties as classical states. Thus
only the collinear B state is possible for $J_3>0$ in the quantum
limit. For $0<J_3<0.5$, the classical ground state has a transition
from a Neel to a collinear B state, with phase transition changing
from $J_2=0.25$ at $J_3=0$ to $J_2=0.5$ at $J_3=0.5$. For $J_3>0.5$,
the classical ground state has a phase transition from a Neel to a
collinear A state. The transition between collinear A to collinear B
state is located at $0.5<J_2<1$ with $J_3=0.5$. These classical
states phase diagram is shown in Fig. \ref{4-class}.

Comparing these classical phase diagram with the quantum phase
diagram in Fig. \ref{4-j3}, the Neel, collinear A and collinear B
states appear in almost the same place in phase diagram, except the
phase boundary varies slightly. The other difference is that there
is a intermediate phase between the Neel and the collinear B state
in the quantum limit. At $J_3=0$, the large $J_2$ phase (purple
region) is the dimer state for quantum case on the XC8-0 cylinder.
In summary, even with an extra parameter $J_3$ included, we still
could not find any trace of a spin liquid state on the honeycomb
lattice with only \textit{XY} interaction.

\section{The transition between $J_1-J_2$ \textit{XY} and Heisenberg model at $J_2=0.3$}

\begin{figure}[htbp]
\begin{center}
\includegraphics*[width=8.5cm,angle=0]{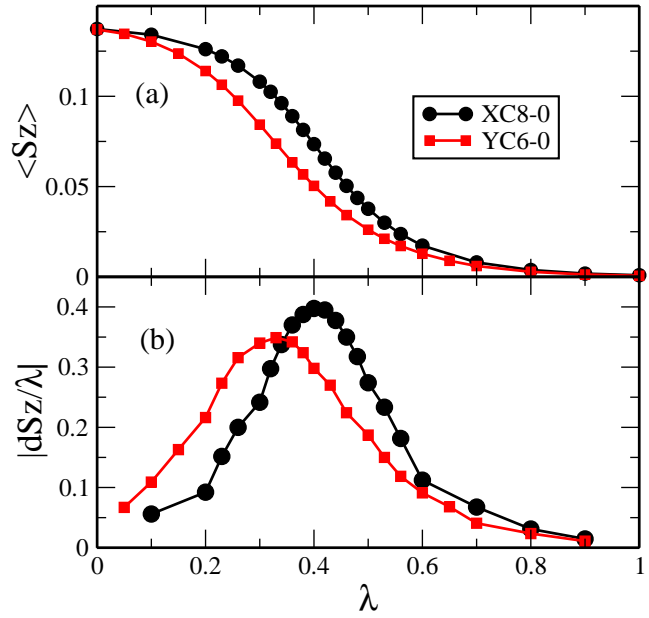}
\caption{(a)The magnetization $\langle S_z \rangle$ in the cylinder
center for XC8-0 and YC6-0 cylinders versus the $S_z$ coupling
$\lambda$ (see Eq. 4) for $J_2=0.3$. (b) The first order derivative
of $\langle S_z \rangle$ versus coupling $\lambda$. The peak
indicates the possible phase transition between $J_1-J_2$
\textit{XY} model and $J_1-J_2$ Heisenberg model for $J_2=0.3$ at
around $0.35\sim0.4$. }\label{xyhei}
\end{center}
\end{figure}

Finally in this section, we briefly discuss the effect when the
$S_z$ interaction is included. The Hamiltonian we considered is:
\begin{eqnarray}
H &=& \sum_{\langle i,j\rangle} J_1(S_i^x S_j^x+ S_i^y S_j^y+\lambda S_i^z
S_j^z) \nonumber \\
&+& \sum_{\langle\langle i,j\rangle\rangle} J_2(S_i^x S_j^x+ S_i^y
S_j^y+\lambda S_i^z S_j^z). \label{eq3}
\end{eqnarray}
For $\lambda=0$, this Hamiltonian is exactly as Eq. (1), it is just
a pure $J_1-J_2$ \textit{XY} model. For $\lambda=1$, the Hamiltonian is the
$J_1-J_2$ Heisenberg model.

The $J_1-J_2$ Heisenberg model has been studied extensively in the
literature for the past few years.
\cite{edh,PVBh,ccmh,sseh,frgh,RVBSLh,spiralh,swh,epsh,vmch} Like the
\textit{XY} model case, the Heisenberg model also has an
intermediate phase. All the references agree that this intermediate
phase has no magnetic order. Initially there has been controversy
about the existence of SL or plaquette valence bond (PVB) order for
this phase. But recent DMRG calculations found that this
intermediate phase has long range PVB order, although there might be
a very narrow parameter regime near the first critical point where
the SL phase is possible. DMRG calculation from Ref. \refcite{cyc}
found the intermediate phase regime between  $0.26 <J_2<0.36$. Ref.
\refcite{dmh1} measured the two phase transition points located at
0.22 and 0.35. Ref. \refcite{dmh2} pointed out the PVB order is
between 0.25 and 0.35, with a narrow regime ($0.22<J_2<0.25$) where
a SL state is possible. However the SL regime might be a finite size
effect, which may disappear on larger systems.

Therefore when tuning the parameter $\lambda$ from 0 to 1 inside the
intermediate phase, there must be a phase transition between Ising
and PVB order, i.e. the $S_zS_z$ interactions terms will disfavor
the Ising order. It would be interesting to see where the phase
transition is. Next we will focus on this transition only at
$J_2=0.3$ for the intermediate phase.

In Fig. \ref{xyhei}, we present the magnetization calculated on
XC8-0 and YC6-0 cylinders with Hamiltonian in Eq. \ref{eq3} for
various $\lambda$ at $J_2=0.3$. We also apply a pinning magnetic
field $h_z$ in both the cylinder edges to favor only one kind of
Ising pattern. For a pure \textit{XY} model with $\lambda=0$,
magnetization in the cylinder center is maximum. However, with
increasing $\lambda$, magnetization decreases faster at $\lambda
\sim 0.4$ to almost zero at the Heisenberg limit with $ \lambda=1$.
The peak of the first order derivative of $\langle S_z \rangle$ can
roughly determine where the magnetization changes fastest. The peak
is located at 0.40 for the XC8-0 cylinder and 0.35 for the YC6-0
cylinder. Thus there may be a possible phase transition around
$0.35\sim 0.40 $. It would be useful to perform similar calculations
on wider cylinders like XC10-0 and XC12-0 in the future. If the
derivative curve becomes sharper and position of the peak does not
change too much, we can determine the phase transition points
confidently in the thermodynamic limit.

A recent paper\cite{xxz} by Li \textit{et. al.} studied the same
Hamiltonian as Eq. \ref{eq3} with the coupled cluster method for the
whole phase diagram in the parameter $J_2-\lambda$ space. They found
that as increasing $S_zS_z$ coupling $\lambda$, the Ising order
phase changes to  a possible SL state for $0.21<J_2<0.28(2)$ and PVB
order for $0.28(2)<J_2<0.38$. At $J_2=0.3$, they found the
transition at $\lambda=0.65$, much larger than
$\lambda=0.35\sim0.40$ from the DMRG calculations. Meanwhile the
boundary between SL and PVB order state has much larger uncertainty.
At the Heisenberg limit, they suggested a larger regime of possible
SL states ($0.21<J_2<0.28(2)$), much larger than the regime claimed
by DMRG\cite{dmh2} ($0.22<J_2<0.25$). Therefore, it is still an
unsolved issue for the Heisenberg limit, whether there would be a
possible SL phase near the first phase transition point. Perhaps
future numerical studies from the tensor network method, which works
directly toward the thermodynamic limit, such as projected entangled
pair states (PEPS)\cite{peps,ipeps1,ipeps2} and multi-scale
entanglement renormalization ansatz (MERA)\cite{mera1,mera2} would
provide some new insights into the possible existence of the SL
phase in the phase diagram.

\section{Summary}

In summary, we have reviewed the study of $J_1-J_2$
antiferromagnetic spin-1/2 \textit{XY} model on the honeycomb
lattice. Instead of a spin-liquid ground state in the intermediate
phase regime for $0.22<J_2/J_1<0.36$, there exists an Ising ordered
phase with a staggered magnetization along the $z$ direction that
does not show any strong finite size effects. Its ground state
energy is much lower than proposed spin-liquid states with a
vanishing topological entanglement entropy.  Thinking about this in
terms of the spin model, it is somewhat puzzling to understand why
this phase appears, since there are no $S^z_iS^z_j$ interaction
terms in the spin Hamiltonian. It seems like only the \textit{XY}
model honeycomb lattice with low coordination number has this exotic
phase. We do not find any Ising ordered phase of the same model on
the square, kagome or triangular lattice. Describing the system
instead as hard-core bosons with frustrated hopping, this Ising
phase is then a Mott insulator with one boson per two-site unit
cell, and the Ising order is then CDW order that breaks the $Z_2$
sublattice symmetry of the unit cell. The on-site hard-core
interaction must induce a first-neighbor repulsion that stabilizes
this CDW order. Thus, although this model unfortunately does not
appear to exhibit a spin-liquid ground state, it exhibits this
somewhat surprising CDW $Z_2$ ordered phase. Although the third
neighbor $J_3$ \textit{XY} interaction or the $S_i^zS_j^z$
interaction disfavor the Ising ordered phase, the system either
evolves into a PVB ordered state or a magnetic ordered state,
instead of a SL state.

\section*{Acknowledgments}

We would like to thank David Huse, Sasha Chernyshev, Leonid 
Glazman, Andreas Laeuchli, Sid Parameswaran, Rajiv Singh, 
Raymond Bishop, Marcos Rigol, Victor Galitski, Tigran
Sedrakyan, Juan Carrasquilla, Hongcheng Jiang, Leon Balents,
Miles Stoudenmire, and Simeng Yan for many helpful
discussions.  This work was supported by NSF Grant No. 
DMR-1161348 (Z.Z., S.R.W.).


\begin{thebibliography}{99}

\bibitem{ARVB}
P. W. Anderson, Mater. Res. Bull. \textbf{8}, 153, (1973).

\bibitem{tri0}
D. A. Huse and U. Elser, Phys. Rev. Lett. \textbf{60}, 253 (1988).

\bibitem{SL}
L. Balents, Nature (London) \textbf{464}, 199 (2010) and references therein.

\bibitem{qdm1}
D. S. Rokhsar and S. A. Kivelson, Phys. Rev. Lett. \textbf{61}, 2376 (1988)

\bibitem{qdm2}
R. Moessner and S. L. Sondhi, Phys. Rev. Lett. \textbf{86}, 1881 (2001)

\bibitem{tori}
A. Y. Kitaev, Ann. Phys. \textbf{303}, 2 (2003)

\bibitem{lwen}
M. A. Levin and X. G. Wen, Phys. Rev. B \textbf{71}, 045110 (2005)

\bibitem{kag1}
S. Yan, D. A. Huse, and S. R. White, Science \textbf{332}, 1173
(2011).

\bibitem{ent1}
A. Kitaev and J. Preskill, Phys. Rev. Lett. \textbf{96},  110404  (2006).

\bibitem{ent2}
M. Levin and X. G. Wen, Phys. Rev. Lett. \textbf{96},  110405  (2006).

\bibitem{ex1}
J. S. Helton, K. Matan, M. P. Shores, E. A. Nytko, B. M. Bartlett,
Y. Yoshida, Y. Takano, A. Suslov, Y. Qiu, J.-H. Chung, D. G. Nocera,
and Y. S. Lee, Phys. Rev. Lett. \textbf{98}, 107204 (2007).

\bibitem{ex2}
M. A. de Vries, K. V. Kamenev, W. A. Kockelmann, J. Sanchez-Benitez,
and A. Harrison, Phys. Rev. Lett. \textbf{100}, 157205 (2008).

\bibitem{ex3}
J. S. Helton, K. Matan, M. P. Shores, E. A. Nytko, B. M. Bartlett,
Y. Qiu, D. G. Nocera, and Y. S. Lee, Phys. Rev. Lett. \textbf{104},
147201 (2010).

\bibitem{ex4}
T. H. Han, J. S. Helton, S. Chu, A. Prodi, D. K. Singh, C. Mazzoli,
P. Muller, D. G. Nocera, and Y. S. Lee, Phys. Rev. B \textbf{83},
100402(R) (2011).

\bibitem{ex5}
T. H. Han, J. S. Helton, S. Chu, D. G. Nocera, J. A. R.-Rivera, C.
Broholm, and Y. S. Lee, Nature (London) \textbf{492}, 406 (2012).

\bibitem{kag2}
S. Depenbrock, I. P. McCulloch, and U. Schollwock, Phys. Rev. Lett.
\textbf{109}, 067201 (2012).

\bibitem{kag3}
H. C. Jiang, Z. H. Wang, and L. Balents, Nat. Phys. \textbf{8},
902 (2012).

\bibitem{dmrg1}
S. R. White, Phys. Rev. Lett. \textbf{69}, 2863 (1992).

\bibitem{dmrg2}
S. R. White, Phys. Rev. B \textbf{48}, 10345 (1993).

\bibitem{rev}
E. M. Stoudenmire and S. R. White, Annu. Rev. Condens. Matter Phys.
\textbf{3}, 111 (2012).

\bibitem{hubhon}
Z. Y. Meng, T. C. Lang, S. Wessel, F. F. Assaad and A.
Muramatsu, Nature \textbf{464}, 847, (2010).

\bibitem{squa1}
H. Jiang, H. Yao and L. Balents, Phys. Rev. B 86, 024424, (2012).

\bibitem{squa2}
L. Wang, Z.-C. Gu, X.-G. Wen and F. Verstraete, arXiv:1112.3331, (unpublished).

\bibitem{hubhon2}
Sandro Sorella, Yuichi Otsuka, Seiji Yunoki, Scientific
Reports \textbf{2}, 992, (2012).

\bibitem{jq}
A. W. Sandvik, Phys. Rev. B \textbf{85}, 134407, (2012).

\bibitem{nosl1}
F. F. Assaad and I. F. Herbut, Phys. Rev. X \textbf{3}, 031010 (2013).

\bibitem{nosl2}
S.-S. Gong, W. Zhu, D. N. Sheng, O. I. Motrunich, M. P. A. Fisher, Phys. Rev. Lett.
\textbf{113}, 027201 (2014).

\bibitem{bose}
C. N. Varney, K. Sun, V. Galitski, and M. Rigol, Phys. Rev. Lett.
\textbf{107}, 077201 (2011).

\bibitem{bo1}
D. N. Sheng, O. I. Motrunich, and M. P. A. Fisher, Phys. Rev. B
\textbf{79}, 205112 (2009).

\bibitem{bo2}
M. S. Block, D. N. Sheng, O. I. Motrunich, and M. P. A. Fisher, Phys.
Rev. Lett. \textbf{106}, 157202 (2011).

\bibitem{bo3}
I. Kimchi, S. A. Parameswaran, A. M. Turner, F. Wang and A.
Vishwanath, Proc. Natl. Acad. Sci. U.S.A. \textbf{110}, 16378 (2013).

\bibitem{mf}
T. A. Sedrakyan, L. I. Glazman, and A. Kamenev, Phys. Rev. B
\textbf{89}, 201112 (2014)

\bibitem{zhu}
Z. Zhu, D. A. Huse, and S. R. White, Phys. Rev. Lett. \textbf{111},
257201 (2013).

\bibitem{cyc}
Z. Zhu, D. A. Huse, and S. R. White, Phys. Rev. Lett. \textbf{110},
127205 (2013).

\bibitem{cla}
A. Di Ciolo, J. Carrasquilla,  F. Becca, M. Rigol, and V. Galitski, Phys. Rev. B
\textbf{89}, 094413 (2014).

\bibitem{sexy}
J. Oitmaa, AND R. R. P. Singh, Phys. Rev. B \textbf{89}, 104423 (2014).

\bibitem{cmxy}
R. F. Bishop, P. H. Y. Li, and C. E. Campbell, Phys. Rev. B \textbf{89},
214413 (2014)

\bibitem{ea}
J. Eisert, M. Cramer, and M. B. Plenio, Rev. Mod. Phys.  \textbf{82}, 277 (2010).

\bibitem{vmc}
J. Carrasquilla, A. Di Ciolo, F. Becca, V. Galitski, and M. Rigol,
Phys. Rev. B \textbf{88}, 241109 (2013).

\bibitem{edh}
A. F. Albuquerque, D. Schwandt, B. Het\'{e}nyi, S. Capponi,
  M. Mambrini and A. M. L\"{a}uchli, Phys. Rev. B \textbf{84}, 024406, (2011).

\bibitem{PVBh}
H. Mosadeq, F. Shahbazi, and S. A. Jafari, J. Phys.: Condens.
  Matter \textbf{23}, 226006, (2011).

\bibitem{ccmh}
D. J. J. Farnell, R. F. Bishop, P. H. Y. Li, J. Richter and
  C. E. Campbell, Phys. Rev. B \textbf{84}, 012403, (2011).

\bibitem{sseh}
J. Oitmaa, and R. R. P. Singh, Phys. Rev. B \textbf{84}, 094424,  (2011).

\bibitem{frgh}
J. Reuther, D. A. Abanin, and R. Thomale, Phys. Rev. B
  \textbf{84}, 014417, (2011).

\bibitem{RVBSLh}
J. B. Fouet, P. Sindzingre, C. Lhuillier, Eur. Phys. J. B.
  \textbf{20}, 241, (2001).

\bibitem{spiralh}
A. Mulder, R. Ganesh, L. Capriotti, and A. Paramekanti, Phys.
  Rev. B \textbf{81}, 214419,  (2010).

\bibitem{swh}
D. C. Cabra, C. A. Lamas, and H. D. Rosales, Mod. Phys. Lett.
  B \textbf{25}, 891,  (2011).

\bibitem{epsh}
F. Mezzacapo and M. Boninsegni, Phys. Rev. B \textbf{85},
  060402,  (2012).

\bibitem{vmch}
B. K. Clark, D. A. Abanin, and S. L. Sondhi, Phys. Rev. Lett.
  \textbf{332}, 1173,  (2011).

\bibitem{dmh1}
R. Ganesh, J. van den Brink, and S. Nishimoto,
Phys. Rev. Lett. \textbf{110} 127203, (2013).

\bibitem{dmh2}
S.-S. Gong, D. N. Sheng, O. I. Motrunich, and M. P. A. Fisher,
Phys. Rev. B \textbf{88} 165138, (2013).

\bibitem{xxz}
P. H. Y. Li, R. F. Bishop, and C. E. Campbell,
Phys. Rev. B \textbf{89} 220408, (2014).

\bibitem{peps}
F. Verstraete, V. Murg and J. Cirac, Adv. Phys. \textbf{57} 143, (2008).

\bibitem{ipeps1}
J. Jordan, R. Orus, G. Vidal, F. Verstraete and J. Cirac,  Phys. Rev. Lett. \textbf{101}
250602, (2008).

\bibitem{ipeps2}
R. Orus and G. Vidal, Phys. Rev. B \textbf{80} 094403, (2009).

\bibitem{mera1}
G. Vidal,  Phys. Rev. Lett. \textbf{99} 220405, (2007).

\bibitem{mera2}
G. Evenbly and G. Vidal,  Phys. Rev. B \textbf{79} 144108, (2009).

\end{thebibliography}
\end{document}